\documentclass[aps]{revtex4}%
\usepackage{amssymb}
\usepackage{amsfonts}
\usepackage{amsmath}
\usepackage{graphicx}%
\providecommand{\U}[1]{\protect\rule{.1in}{.1in}}
\begin{document}
\preprint{ }
\title[Optical response of an interacting polaron gas]{Optical response of an interacting polaron gas in strongly polar crystals}
\author{S. N. Klimin}
\author{J. Tempere}
\thanks{Also at Lyman Laboratory of Physics, Harvard University, Cambridge, MA 02138, USA.}
\affiliation{Theorie van Kwantumsystemen en Complexe Systemen (TQC), Universiteit
Antwerpen, Universiteitsplein 1, B-2610 Antwerpen, Belgium}
\author{J. T. Devreese}
\affiliation{Theorie van Kwantumsystemen en Complexe Systemen (TQC), Universiteit
Antwerpen, Universiteitsplein 1, B-2610 Antwerpen, Belgium}
\author{C. Franchini}
\author{G. Kresse}
\affiliation{Faculty of Physics, Computational Materials Physics, University of Vienna,
Vienna A-1090, Austria}

\begin{abstract}
Optical conductivity of an interacting polaron gas is calculated within an
extended random phase approximation which takes into account mixing of
collective excitations of the electron gas with LO phonons. This mixing is
important for the optical response of strongly polar crystals where the static
dielectric constant is rather high: strontium titanate is the case. The
present calculation sheds light on unexplained features of experimentally
observed optical conductivity spectra in $n$-doped SrTiO$_{3}$. These features
appear to be due to dynamic screening of the electron-electron interaction by
polar optical phonons and hence do not require additional mechanisms for the explanation.

\end{abstract}
\date{\today}
\maketitle

\section{Introduction}

Polaron manifestations in the optical response of polar crystals, such as
complex oxides and high-$T_{c}$ superconductors are the subject of intense
investigations \cite{Lupi1999,calva0,calva1b,QQ4,zhang,Homes1997}. Several
features in the infrared optical absorption spectra of cuprates have been
associated with large polarons \cite{jt1} or with a mixture of large and small
(bi)polarons \cite{emin}. The analysis in those papers was performed using a
single-polaron picture, so that the density (doping) dependence of the
optical-absorption spectra could not be studied.

The many-body theory of the optical absorption of a gas of interacting
polarons \cite{TD2001,K2010} allows one to study the density dependence of the
optical-absorption spectra. The calculation \cite{TD2001} was performed in the
single-branch approximation for LO phonons. In Ref. \cite{K2010}, the optical
conductivity of $n$-doped SrTiO$_{3}$ was calculated accounting for the
electron-phonon interaction with multiple LO-phonon branches. The calculations
of the optical conductivity of a weak-coupling polaron gas \cite{TD2001,K2010}
compare fairly well with the experimental data \cite{Lupi1999,VDM-PRL2008} and
therefore confirm the contribution of large polaron in the optical response.
Strontium titanate represents an especially interesting case due to its unique
features, particularly a high static dielectric constant at low temperatures
and essentially nonparabolic shape of the conduction band which consists of
three subbands. This requires a treatment of the optical conductivity beyond
the frequently used lowest-order perturbation approximation.

The first-principle methods are powerful for the theoretical study of both
equilibrium and response properties of polarons. At present, \emph{ab initio}
calculations of the polaron band energies are developed, for example, in Refs.
\cite{Franchini1,Giustino,Sio1,Sio2}. In the present work, we consider a
complementary semianalytic approach, which has its own advantages. First, it
is much less time- and memory-consuming for computation. Second, more
important, it allows sometimes a clear physical interpretation of features of
obtained spectra, as they follow from a used model.

The present work is focused on the many-polaron optical response in strongly
polar crystals like SrTiO$_{3}$. The strong polarity means that the ratio of
the static and high-frequency dielectric constant is large: $\varepsilon
_{0}/\varepsilon_{\infty}\gg1$. This does not necessarily lead to a high
electron-phonon coupling constant: in strontium titanate the effective
$\alpha\approx2$ as determined in \cite{K2010}. Even in this moderate-coupling
case, electron collective excitations (attributed to plasmons only in the
long-wavelength limit) are substantially mixed with LO phonons \cite{DL1} and
therefore can result in a non-trivial spectrum of the optical conductivity.
Therefore the optical conductivity in a many-polaron gas is considered
accounting for mixing of electron and phonon collective excitations in the
total dielectric function of the electron-phonon system, and taking into
account multi-subband structure of the conductivity band. The method is
applied to $n$-doped strontium titanate.

\section{Many-polaron optical conductivity}

We consider an electron-phonon system with the Hamiltonian in the momentum
representation:%
\begin{align}
H  &  =\sum_{\lambda}\sum_{\mathbf{k}}\varepsilon_{\lambda}\left(
\mathbf{k}\right)  \sum_{\sigma=\pm1/2}a_{\mathbf{k},\sigma,\lambda}^{\dag
}a_{\mathbf{k},\sigma,\lambda}+\sum_{\mathbf{q},j}\hbar\omega_{\mathbf{q}%
,j}\left(  b_{\mathbf{q},j}^{\dag}b_{\mathbf{q},j}+\frac{1}{2}\right)
\nonumber\\
&  +\sum_{\mathbf{q},j}V_{\mathbf{q},j}\left(  b_{\mathbf{q},j}+b_{-\mathbf{q}%
,j}^{\dag}\right)  \sum_{\lambda}\sum_{\mathbf{k}}\sum_{\sigma=\pm
1/2}a_{\mathbf{k}+\mathbf{q},\sigma,\lambda}^{\dag}a_{\mathbf{k}%
,\sigma,\lambda}, \label{Htot}%
\end{align}
where $\varepsilon_{\lambda}\left(  \mathbf{k}\right)  $ is the electron
energy with the momentum $\hbar\mathbf{k}$ in the $\lambda$-th subband of the
conduction band, $a_{\mathbf{k},\sigma,\lambda}^{\dag}$ and $a_{\mathbf{k}%
,\sigma,\lambda}$ are, respectively, creation and annihilation fermionic
operators for an electron with the spin projection $\sigma$, $\omega
_{\mathbf{q},j}$ is the phonon frequency for the momentum $\hbar\mathbf{q}$
and the phonon branch $j$, $b_{\mathbf{q},j}^{\dag}$ and $b_{\mathbf{q},j}$
are, respectively, phonon creation and annihilation operators. The
electron-phonon interaction amplitudes $V_{\mathbf{q},j}$ are used here
neglecting their possible dependence on the electron momentum and the subband
number. It can be non-negligible when high-energy electrons bring an important
contribution to the many-polaron response, what is beyond the scope of the
present treatment.

For the many-polaron optical response, we start the Kubo formula,%
\begin{align}
\sigma_{xx}\left(  \Omega\right)   &  =\frac{1}{Vz}\left[  \frac{1}{\hbar}%
\int_{0}^{\infty}dt~e^{izt}\left\langle \left[  J_{x}\left(  t\right)
,J_{x}\left(  0\right)  \right]  \right\rangle +i\mathcal{Z}\right]
\label{s}\\
&  \left(  \beta=\frac{1}{k_{B}T},\;z=\Omega+i\delta,\;\delta\rightarrow
+0\right) \nonumber
\end{align}
where $V$ is the system volume, $e$ is the electronic charge, and the constant
$\mathcal{Z}$ is determined by the integral:%
\begin{equation}
\mathcal{Z}=\frac{1}{\hbar}\int_{0}^{\hbar\beta}d\tau\left\langle J_{x}\left(
\tau\right)  J_{x}\left(  0\right)  \right\rangle , \label{Z}%
\end{equation}
and $J_{x}$ is the current operator determined by:%
\begin{equation}
J_{x}=-ev_{x}=-\frac{e}{\hbar}\sum_{\lambda}\sum_{\mathbf{k}}\sum_{\sigma
=\pm1/2}\frac{\partial\varepsilon_{\lambda}\left(  \mathbf{k}\right)
}{\partial k_{x}}a_{\mathbf{k},\sigma,\lambda}^{\dag}a_{\mathbf{k}%
,\sigma,\lambda}. \label{J}%
\end{equation}
The constant $\mathcal{Z}$ can be calculated explicitly using the commutation
relations for operators. The result is%
\begin{equation}
\mathcal{Z}=\frac{e^{2}}{\hbar^{2}}\sum_{\lambda}\sum_{\mathbf{k}}\sum
_{\sigma}\frac{\partial^{2}\varepsilon_{\lambda}\left(  \mathbf{k}\right)
}{\partial k_{x}^{2}}f_{\mathbf{k},\sigma,\lambda}, \label{Z1}%
\end{equation}
with the distribution function of the electrons,%
\begin{equation}
f_{\mathbf{k},\sigma}=\left\langle a_{\mathbf{k},\sigma,\lambda}^{\dag
}a_{\mathbf{k},\sigma,\lambda}\right\rangle . \label{nk}%
\end{equation}
Next, we perform twice the integration by parts in the integral over time in
(\ref{s}) and introduce the force operator,%
\begin{equation}
\frac{\partial J_{x}\left(  t\right)  }{\partial t}\equiv e\mathcal{F}%
_{x}\left(  t\right)  . \label{F}%
\end{equation}
which is explicitly given by the expression:%
\begin{equation}
\mathcal{F}_{x}=\frac{i}{\hbar}\sum_{\mathbf{q},j}V_{\mathbf{q},j}\left(
b_{\mathbf{q},j}+b_{-\mathbf{q},j}^{\dag}\right)  \hat{B}_{\mathbf{q}},
\label{F2}%
\end{equation}
with%
\begin{equation}
\hat{B}_{\mathbf{q}}\equiv\frac{1}{\hbar}\sum_{\lambda}\sum_{\mathbf{k}}%
\sum_{\sigma}\left(  \frac{\partial\varepsilon_{\lambda}\left(  \mathbf{k}%
+\mathbf{q}\right)  }{\partial k_{x}}-\frac{\partial\varepsilon_{\lambda
}\left(  \mathbf{k}\right)  }{\partial k_{x}}\right)  a_{\mathbf{k}%
,\sigma,\lambda}^{\dag}a_{\mathbf{k},\sigma,\lambda} \label{Bq}%
\end{equation}
After the twice integration by parts, the Kubo formula is equivalently
rewritten through the force-force correlation function,%
\begin{equation}
\sigma_{xx}\left(  \Omega\right)  =\frac{e^{2}}{\hbar V\left(  \Omega
+i\delta\right)  ^{3}}\int_{0}^{\infty}dt~e^{-\delta t}\left(  e^{i\Omega
t}-1\right)  \left\langle \left[  \mathcal{F}_{x}\left(  t\right)
,\mathcal{F}_{x}\left(  0\right)  \right]  \right\rangle +\frac{i}{V}%
\frac{\mathcal{Z}}{\Omega+i\delta}. \label{sig}%
\end{equation}

Next, we consider the weak-coupling regime. The weak-coupling optical
conductivity can be expressed in the memory-function form, as, e.~g., in Refs.
\cite{PD1983,DF2014}:%
\begin{equation}
\sigma_{xx}\left(  \Omega\right)  =\frac{i}{V}\frac{\mathcal{Z}}%
{\Omega+i\delta-\chi\left(  \Omega\right)  /\left(  \Omega+i\delta\right)  }.
\end{equation}
where the memory function $\chi\left(  \Omega\right)  $ is:%
\begin{equation}
\chi\left(  \Omega\right)  =-\frac{i}{\mathcal{Z}}\frac{e^{2}}{\hbar}\int
_{0}^{\infty}dt~e^{-\delta t}\left(  e^{i\Omega t}-1\right)  \left\langle
\left[  \mathcal{F}_{x}\left(  t\right)  ,\mathcal{F}_{x}\left(  0\right)
\right]  \right\rangle _{0}. \label{hi}%
\end{equation}
Here, the averaging $\left\langle \ldots\right\rangle _{0}$ is performed with
the Hamiltonian of interacting electrons neglecting the electron-phonon interaction.

We can transform the memory function in an explicitly tractable expression
substituting (\ref{F2}) and (\ref{Bq}) in (\ref{hi}). Thus we obtain the
resulting memory function:%
\begin{equation}
\chi\left(  \Omega\right)  =\frac{2e^{2}}{\hbar^{3}\mathcal{Z}}\sum
_{\mathbf{q},j}\left\vert V_{\mathbf{q},j}\right\vert ^{2}\int_{0}^{\infty
}dt~e^{-\delta t}\left(  e^{i\Omega t}-1\right)  \operatorname{Im}\left[
T^{\ast}\left(  \omega_{\mathbf{q},j},t\right)  \left\langle \hat
{B}_{\mathbf{q}}\left(  t\right)  \hat{B}_{\mathbf{q}}^{\dag}\right\rangle
_{0}\right]  .
\end{equation}
where $T\left(  \omega_{\mathbf{q},j},t\right)  $ is the phonon Green's
function:%
\begin{equation}
T\left(  \omega_{\mathbf{q},j},t\right)  =\left(  1+\bar{n}_{\mathbf{q}%
,j}\right)  e^{i\omega_{\mathbf{q},j}t}+\bar{n}_{\mathbf{q},j}e^{-i\omega
_{\mathbf{q},j}t}. \label{Tf}%
\end{equation}
and $\bar{n}_{\mathbf{q},j}$ is the Bose distribution of phonons:%
\begin{equation}
\bar{n}_{\mathbf{q},j}=\frac{1}{e^{\beta\hbar\omega_{\mathbf{q},j}}-1}.
\end{equation}
The $f$-sum rule for the optical conductivity reads:
\begin{equation}
\int_{-\infty}^{\infty}\operatorname{Re}\sigma_{xx}\left(  \Omega\right)
d\Omega=\frac{\pi}{2}\frac{\mathcal{Z}}{V}. \label{SR}%
\end{equation}
In the general case the constant $\mathcal{Z}$ can be different from the value
$\frac{e^{2}N_{e}}{m_{b}}$ obtained in Ref. \cite{DLR1977}, which follows from
(\ref{Z1}) for a quadratic dispersion.

\section{Semianalytic approximations}

The optical conductivity of an interacting polaron gas is calculated here
within the extended random phase approximation (RPA), as described below. The
memory function $\chi\left(  \Omega\right)  $ can be expressed through the
polarization function of the electron gas for sufficiently small $\mathbf{q}$.
Thus the RPA can be applied under the assumption that the long-wavelength
phonons bring a dominating contribution to the polaron optical response. Thus
the memory function for the optical conductivity is approximated by the
expression:
\begin{equation}
\chi\left(  \Omega\right)  =\frac{2e^{2}}{3\hbar m_{b}^{2}\mathcal{Z}}%
\sum_{\mathbf{q},j}\left\vert V_{\mathbf{q},j}\right\vert ^{2}q^{2}\int
_{0}^{\infty}dt~e^{-\delta t}\left(  e^{i\Omega t}-1\right)  \operatorname{Im}%
\left[  T^{\ast}\left(  \omega_{\mathbf{q},j},t\right)  \left\langle
\rho_{\mathbf{q}}\left(  t\right)  \rho_{\mathbf{q}}^{\dag}\right\rangle
_{0}\right]  , \label{xi}%
\end{equation}
where $m_{b}$ is the band mass, and $\rho_{\mathbf{q}}=\sum_{\mathbf{k}%
,\sigma,\lambda}a_{\mathbf{k}+\mathbf{q},\sigma,\lambda}^{\dag}a_{\mathbf{k}%
,\sigma,\lambda}$ is the Fourier component of the electron density.

Here, we use the Fr\"{o}hlich interaction amplitudes with the partial coupling
constants $\alpha_{j}$ for the $j$-th phonon branch:%
\begin{equation}
V_{\mathbf{q},j}=\frac{\hbar\omega_{L,j}}{q}\left(  \frac{4\pi\alpha_{j}}%
{V}\right)  ^{1/2}\left(  \frac{\hbar}{2m_{b}\omega_{L,j}}\right)  ^{1/4}.
\label{V2}%
\end{equation}
In terms of Green's functions, the memory function (\ref{xi}) takes the form:%
\begin{align}
\chi\left(  \Omega\right)   &  =\sum_{j}\frac{\alpha_{j}\hbar\omega_{L,j}%
^{2}e^{2}}{6\pi^{2}m_{b}^{2}\mathcal{Z}}\left(  \frac{\hbar}{2m_{b}%
\omega_{L,j}}\right)  ^{1/2}\nonumber\\
&  \times\int d\mathbf{q}\left\{  \mathcal{G}\left(  \mathbf{q},\Omega
-\omega_{L,j}\right)  +\mathcal{G}^{\ast}\left(  \mathbf{q},-\Omega
-\omega_{L,j}\right)  -\mathcal{G}\left(  \mathbf{q},-\omega_{L,j}\right)
-\mathcal{G}^{\ast}\left(  \mathbf{q},-\omega_{L,j}\right)  \right.
\nonumber\\
&  +\frac{1}{e^{\beta\hbar\omega_{L,j}}-1}\left[  G^{R}\left(  \mathbf{q}%
,\Omega-\omega_{L,j}\right)  +\left(  G^{R}\left(  \mathbf{q},-\Omega
-\omega_{L,j}\right)  \right)  ^{\ast}\right. \nonumber\\
&  \left.  \left.  -G^{R}\left(  \mathbf{q},-\omega_{L,j}\right)  -\left(
G^{R}\left(  \mathbf{q},-\omega_{L,j}\right)  \right)  ^{\ast}\right]
\right\}  . \label{MF}%
\end{align}
with
\begin{align}
\mathcal{G}\left(  \mathbf{q},\Omega\right)   &  \equiv-i\int_{0}^{\infty
}e^{i\Omega t}\left\langle \rho_{\mathbf{q}}\left(  t\right)  \rho
_{-\mathbf{q}}\left(  0\right)  \right\rangle _{0}~dt,\label{GF1}\\
G^{R}\left(  \mathbf{q},\Omega\right)   &  \equiv-i\int_{0}^{\infty}e^{i\Omega
t}\left\langle \left[  \rho_{\mathbf{q}}\left(  t\right)  ,\rho_{-\mathbf{q}%
}\left(  0\right)  \right]  \right\rangle _{0}~dt. \label{GF2}%
\end{align}
In Ref. \cite{K2010}, the Green's functions were calculated within the RPA for
an electron gas. Here, we apply the RPA extended for an interacting
electron-phonon system, which leads to the formula structurally similar to
that obtained within RPA, but with the different (momentum and frequency
dependent) electron-electron interaction matrix element:%
\begin{equation}
\frac{4\pi e^{2}}{\varepsilon_{\infty}q^{2}}\rightarrow\frac{4\pi e^{2}%
}{\varepsilon_{L}\left(  \mathbf{q},\omega\right)  q^{2}}, \label{corr}%
\end{equation}
with the dielectric function of the lattice $\varepsilon_{L}\left(
\mathbf{q},\omega\right)  $, which describes the dynamic lattice polarization.
In the present calculation, we use the model of independent oscillators
\cite{T1972,mmc3} which correspond to the LO and TO phonon modes:%
\begin{equation}
\varepsilon_{L}\left(  \mathbf{q},\Omega\right)  =\varepsilon_{\infty}%
\prod_{j=1}^{n}\left(  \frac{\Omega^{2}-\omega_{L,j}^{2}\left(  \mathbf{q}%
\right)  }{\Omega^{2}-\omega_{T,j}^{2}\left(  \mathbf{q}\right)  }\right)  .
\label{DF}%
\end{equation}
The extended RPA thus takes into account the dynamic screening of the Coulomb
electron-electron interaction by the lattice polarization. The resulting
retarded density-density Green's function is:%
\begin{equation}
G^{R}\left(  \mathbf{q},\Omega\right)  =\frac{\hbar VP^{\left(  1\right)
}\left(  \mathbf{q},\Omega\right)  }{1-\frac{4\pi e^{2}}{\varepsilon
_{L}\left(  \mathbf{q},\Omega\right)  q^{2}}P^{\left(  1\right)  }\left(
\mathbf{q},\Omega\right)  }, \label{GH}%
\end{equation}
where $P^{\left(  1\right)  }\left(  \mathbf{q},\Omega\right)  $ is the
Lindhard polarization function,%
\begin{equation}
P^{\left(  1\right)  }\left(  \mathbf{q},\Omega\right)  =\frac{1}{V}%
\sum_{\mathbf{k},\sigma,\lambda}\frac{f_{F}\left(  \varepsilon_{\lambda
}\left(  \mathbf{k}+\mathbf{q}\right)  -\mu_{\lambda}\right)  -f_{F}\left(
\varepsilon_{\lambda}\left(  \mathbf{k}\right)  -\mu_{\lambda}\right)  }%
{\hbar\Omega+\varepsilon_{\lambda}\left(  \mathbf{k}+\mathbf{q}\right)
-\varepsilon_{\lambda}\left(  \mathbf{k}\right)  +i0^{+}} \label{P1a}%
\end{equation}
with the Fermi distribution function $f_{F}\left(  \varepsilon\right)  $:%
\begin{equation}
f_{F}\left(  \varepsilon\right)  =\frac{1}{e^{\beta\left(  \varepsilon
-\mu\right)  }+1}.
\end{equation}
The function $\mathcal{G}\left(  \mathbf{q},\Omega\right)  $ is obtained from
$G^{R}\left(  \mathbf{q},\Omega\right)  $ using the analytic identity,%
\begin{equation}
\left(  1-e^{-\beta\hbar\Omega}\right)  \operatorname{Im}\mathcal{G}\left(
\mathbf{q},\Omega\right)  =\operatorname{Im}G^{R}\left(  \mathbf{q}%
,\Omega\right)  , \label{pr1}%
\end{equation}
and then the Kramers-Kronig dispersion relation for $\operatorname{Re}%
\mathcal{G}\left(  \mathbf{q},\Omega\right)  $.

For a comparison with experiment, the Green's functions are calculated here
accounting for damping within the Mermin-Lindhard approach
\cite{Mermin,Arkhipov,Ropke,Barriga} (where the damping is introduced in such
a way to conserve the local electron number). This leads to the modification
of the retarded density-density Green's function as follows:%
\begin{equation}
G^{R}\left(  \mathbf{q},\Omega\right)  \rightarrow G_{M}^{R}\left(
\mathbf{q},\Omega,\gamma\right)  =G^{R}\left(  \mathbf{q},\Omega
+i\gamma\right)  \frac{\Omega+i\gamma}{\Omega+i\gamma\frac{G^{R}\left(
\mathbf{q},\Omega+i\gamma\right)  }{G^{R}\left(  \mathbf{q},0\right)  }}.
\label{GRM}%
\end{equation}
where $\gamma$ is the phenomenological damping factor. The values of $\gamma$
found in the literature are of the order of the Fermi energy of electrons
\cite{Barriga2,Morawetz}. Here, $\gamma$ is a fitting parameter of the same
order of magnitude (the \emph{only} fitting parameter which is in fact used
here). The calculation is performed with the values $\gamma=1.2\varepsilon
_{F,1}$ for $T=7%
%TCIMACRO{\unit{K}}%
%BeginExpansion
\operatorname{K}%
%EndExpansion
$ and $\gamma=2\varepsilon_{F,1}$ for $T=300%
%TCIMACRO{\unit{K}}%
%BeginExpansion
\operatorname{K}%
%EndExpansion
$. It should be noted that the results appear to be only slightly sensitive to
chosen values of $\gamma$.

The calculation of the Green's functions for non-parabolic bands requires
knowledge of overlap integrals \cite{Yamaguchi} for the Coulomb and
electron-phonon interactions, which is not yet reliably known and needs a
microscopic calculation. In order to simplify the computation keeping main
features of the non-parabolic band dispersion, we perform two approximations.

First, we apply the density-of-states approach already successfully used in
Ref. \cite{JSNM-2019}. The approximation consists in the replacement of the
true band energy $\varepsilon_{\lambda}\left(  \mathbf{k}\right)  $ by the
model isotropic band energy $\varepsilon_{\lambda}\left(  k\right)  $ which
provides \emph{the same density of states} as that for the true band energy
$\varepsilon_{\lambda}\left(  \mathbf{k}\right)  $. The density of states
$\nu_{\lambda}\left(  E\right)  $ in the $\lambda$-th subband of the
conductivity band is determined using the carrier density:%
\begin{equation}
n_{\lambda}=\frac{1}{4\pi^{3}}\int d\mathbf{k}~f\left(  \varepsilon_{\lambda
}\left(  \mathbf{k}\right)  -\mu\right)  =\int_{\varepsilon_{\lambda,\min}%
}^{\varepsilon_{\lambda,\max}}f\left(  E-\mu_{\lambda}\right)  \nu_{\lambda
}\left(  E\right)  dE. \label{n}%
\end{equation}
where $\varepsilon_{\lambda,\min}=\left.  \varepsilon_{\lambda}\left(
\mathbf{k}\right)  \right\vert _{k=0}$. The model isotropic band energy
dispersion is determined through the function%
\begin{equation}
k_{\lambda}\left(  E\right)  =\left(  3\pi^{2}\int_{\varepsilon_{\lambda,\min
}}^{E}\nu_{\lambda}\left(  \varepsilon\right)  d\varepsilon\right)  ^{1/3},
\label{kE}%
\end{equation}
so that $\varepsilon_{\lambda}\left(  k\right)  $ is the inverse function to
this $k_{\lambda}\left(  E\right)  $.

Second, $\varepsilon_{\lambda}\left(  k\right)  $ appears to be approximately
parabolic in a rather wide range of the momentum. Therefore we assume the
parabolic conduction band for the calculation of Green's functions but with
the density-of-states effective masses $m_{\lambda}$ determined through the
density of states from the condition that the low-momentum expansion of the
polarization function $P^{\left(  1\right)  }\left(  \mathbf{q},\Omega\right)
$ with the dispersion $\varepsilon_{\lambda}\left(  k\right)  $ coincides with
that for a parabolic band dispersion with the mass $m_{\lambda}$. This gives
us the expression:
\begin{equation}
m_{\lambda}=3\pi^{4}n_{\lambda}\left(  \int_{\varepsilon_{\lambda,\min}%
}^{\varepsilon_{\lambda,\max}}\frac{\beta e^{\beta\left(  E-\mu\right)  }%
}{\left(  e^{\beta\left(  E-\mu\right)  }+1\right)  ^{2}}\frac{k_{\lambda}%
^{4}\left(  E\right)  }{\nu_{\lambda}\left(  E\right)  }dE\right)  ^{-1}.
\label{mla1}%
\end{equation}
In the zero-temperature limit, $m_{\lambda}$ is analytically expressed through
the density of states at the Fermi energy $\varepsilon_{F,\lambda}$:%
\begin{equation}
m_{\lambda}=\pi^{2}\frac{\nu_{\lambda}\left(  \varepsilon_{F,\lambda}\right)
}{k_{F,\lambda}}. \label{mla}%
\end{equation}

\section{Application to SrTiO$_{3}$}

The approach described above is focused mainly on crystals with a high ratio
$\varepsilon_{0}/\varepsilon_{\infty}$ like strontium titanate, where it can
reveal specific features due to their high polarizability. In the previous
treatment of the many-polaron optical conductivity in doped SrTiO$_{3}$
\cite{K2010}, the pronounced peak for $\hbar\Omega\sim130$ meV at a relatively
low temperature remains unexplained. It was suggested in \cite{K2010} that it
might be provided by other (non-polaron) mechanisms, for example, the
small-polaron and mixed-polaron \cite{Eagles} channels for the optical
response. As we show below, additional mechanisms are not necessary for the
explanation of this 130-meV feature.

\begin{figure}
[tbh]
\begin{center}
\includegraphics[
height=5.0025in,
width=4.01in
]%
{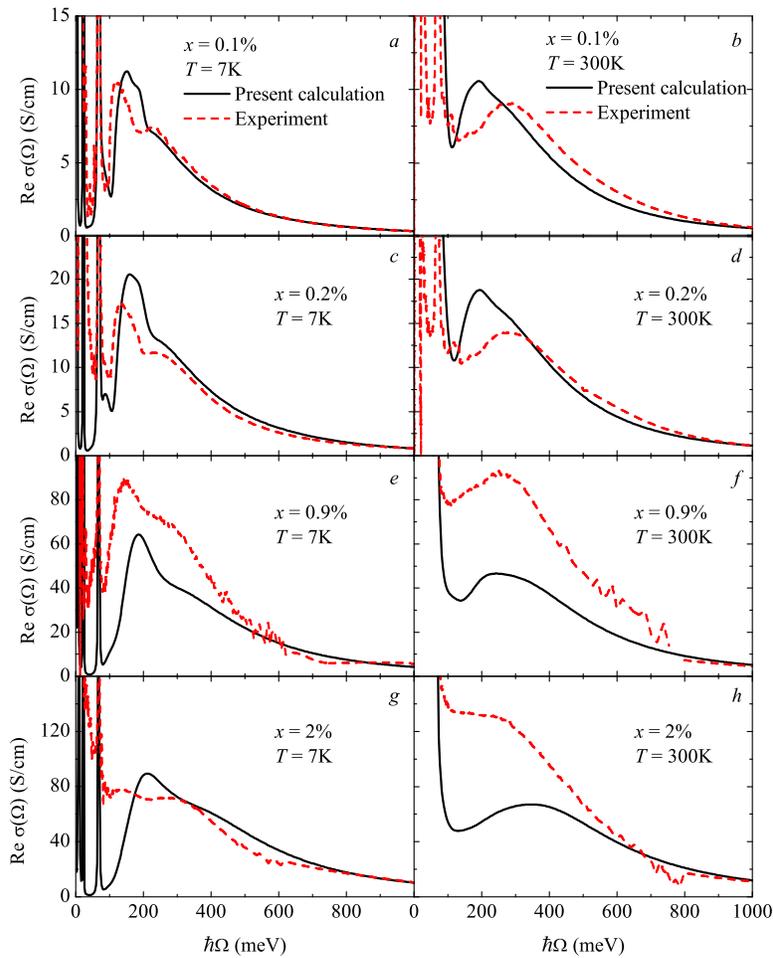}%
\caption{Many-polaron optical conductivity of $n$-doped SrTiO$_{3}$ in the
mid-infrared frequency range for several values of the doping $x$ and two
temperatures (the parameters are indicated in the figure), corresponding to
the experimental conditions of Ref. \cite{VDM-PRL2008}. The calculated spectra
(solid curves) are compared with the experimental data (dashed curves).}%
\label{Figure1}%
\end{center}
\end{figure}

In the computation, the following set of electron band and phonon material
parameters is used. The conduction band shape is simulated by the analytic
tight-binding fit as described in Ref. \cite{VDM2011} and here in Appendix.
The optimal values for this analytic approximation are the diagonal matrix
elements $t_{\delta},t_{\pi}$ corresponding to the recent results of the
microscopic calculation \cite{GW2018} using the GW method \cite{Hedin}:
$t_{\delta}=54.2$ meV, $t_{\pi}=490.9$ meV, and the band splitting parameters
from Ref. \cite{VDM2011} $\xi=18.8$ meV, $D=2.2$ meV. The optical-phonon
energies at the Brillouin zone center of SrTiO$_{3}$ are taken from the
experimental data of Ref. \cite{VDM-PRL2008}, the same as described in Ref.
\cite{K2010}. Also the direct TO-phonon optical response has been included in
the figure in the same way as in Ref. \cite{K2010}. It is represented by sharp
peaks in the low-energy part of the optical conductivity spectrum in Fig.
\ref{Figure1}.%

%TCIMACRO{\FRAME{ftbhFU}{4.01in}{5.0025in}{0pt}{\Qcb{Many-polaron optical
%conductivity of $n$-doped SrTiO$_{3}$ in the mid-infrared frequency range for
%several values of the doping $x$ and two temperatures (the parameters are
%indicated in the figure), corresponding to the experimental conditions of Ref.
%\cite{VDM-PRL2008}. The calculated spectra (solid curves) are compared with
%the experimental data (dashed curves).}}{\Qlb{Figure1}}{figure1.eps}%
%maintain-aspect-ratio TRUE;  display "ICON";  valid_file "F";  width 4.01in;
%height 5.0025in;  depth 0pt;  original-width 5.6086in;
%original-height 7.0061in;  cropleft "0";  croptop "1";  cropright "1";
%cropbottom "0";  filename 'Figure1.eps';file-properties "XNPEU";}} }%
%BeginExpansion
%EndExpansion

The calculated many-polaron optical conductivity in SrTiO$_{3}$ is compared
with the experimental data of Ref. \cite{VDM-PRL2008} for two temperatures:
$T=7%
%TCIMACRO{\unit{K}}%
%BeginExpansion
\operatorname{K}%
%EndExpansion
$ and $T=300%
%TCIMACRO{\unit{K}}%
%BeginExpansion
\operatorname{K}%
%EndExpansion
$ and for several values of the carrier concentration. As can be seen from the
low-temperature results shown in the left-hand panels of the figure, the
130-meV peak and the dip at $\hbar\Omega\sim200$ meV (corresponding to twice
the highest-energy LO phonon mode ion strontium titanate) experimentally
observed at the low temperature and at relatively low concentrations are
fairly revealed in the calculated spectra of the optical conductivity.

The obtained expression (\ref{GH}) for the retarded density-density Green's
function gives us a transparent explanation of the shape of the optical
conductivity spectrum, which is more complicated than in the absence of the
dynamic screening. The Green's functions enter the memory function (\ref{MF})
with the arguments $\left(  \pm\Omega-\omega_{L,j}\right)  $. Therefore the
dynamically screened electron-electron interaction matrix element (\ref{corr})
contains poles, in particular, at $\Omega=2\omega_{L,j}\left(  \mathbf{q}%
\right)  $, which result in dips of the optical conductivity at these
frequencies The most significant contribution to the dips comes from the
highest LO phonon energy $\left.  \hbar\omega_{L,3}\right\vert _{q=0}%
\approx98$ meV. This feature is visible in both the measured and calculated
optical conductivity spectra. The part of the spectrum below $2\omega_{L,3}$
constitutes the aforesaid 130-meV peak. The other part of the spectrum, above
$2\omega_{L,3}$, contains the \textquotedblleft
plasmon-phonon\textquotedblright\ peak provided by the response due to
undamped plasmons \cite{TD2001,K2010}.

\section{Conclusions}

In the present work, we revisit the optical response of a polaron gas in
complex polar crystals using the random phase approximation extended for an
interacting electron-phonon system. This extension results in a modified
many-polaron optical conductivity with an effective electron-electron
interaction accounting for the dynamic screening by LO phonons. For a more
realistic calculation relevant for comparison with experiment for strontium
titanate, the phonon dielectric function contains several optical phonon modes
which actually present in SrTiO$_{3}$.

A distinctive low-frequency peak of the many-polaron optical conductivity in a
polar medium appears when a crystal is highly polar, $\varepsilon
_{0}/\varepsilon_{\infty}\gg1$, which is realized in strontium titanate. As
can be seen from the obtained spectra of the optical conductivity, the dynamic
screening indeed, as expected, leads to an appearance of this peak which is
close to the experimental \textquotedblleft130-meV feature\textquotedblright%
,\ except for the highest available density. Moreover, its width and shape
asymmetry are remarkably similar to those of the experimental peak, including
even fine details such a small kink at the shoulder above the maximum. Also
the whole shape of the spectrum at least for the two lower densities is
similar to the experimental results, containing both the low-frequency peak
and the \textquotedblleft plasmon-phonon\textquotedblright\ peak due to
undamped plasmons. This similarity makes the dynamic screening mechanism for
the low-frequency peak convincing.

There is also a remarkable agreement between the present theory and the
experimental results \cite{VDM-PRL2008} in what concerns the high-frequency
dependence of the optical conductivity (in the range $\hbar\Omega\sim1%
%TCIMACRO{\unit{eV}}%
%BeginExpansion
\operatorname{eV}%
%EndExpansion
$), achieved without adjustment, using reliable material parameters known from
literature. This agreement is in line with experimentally substantiated
conclusion \cite{Meevasana} that polarons in SrTiO$_{3}$ are large rather than small.

\begin{acknowledgments}
We thank D. van der Marel for the experimental data on the optical absorrption
of SrTiO$_{3}$. This work has been supported by the joint FWO-FWF project
POLOX (Grant No. I 2460-N36).
\end{acknowledgments}

\appendix

\section{Analytic model for the conductivity band in SrTiO$_{3}$}

For the calculation of the many-polaron response, it is useful to simulate
numerical data for the band structure by an analytic expression. Here, we
treat the tight-binding expression similarly to Refs. \cite{VDM-PRL2008,K2010}%
. In these works, the matrix Hamiltonian is used for an analytic fit of the
band dispersion law:%
\begin{equation}
H=4\left(
\begin{array}
[c]{ccc}%
\varepsilon_{1}\left(  \mathbf{k}\right)  & 0 & 0\\
0 & \varepsilon_{2}\left(  \mathbf{k}\right)  & 0\\
0 & 0 & \varepsilon_{3}\left(  \mathbf{k}\right)
\end{array}
\right)  +\frac{1}{2}W, \label{H}%
\end{equation}
with the energies%
\begin{align}
\varepsilon_{1}  &  =t_{\delta}\sin^{2}\left(  \frac{a_{0}k_{x}}{2}\right)
+t_{\pi}\sin^{2}\left(  \frac{a_{0}k_{y}}{2}\right)  +t_{\pi}\sin^{2}\left(
\frac{a_{0}k_{z}}{2}\right)  ,\nonumber\\
\varepsilon_{2}  &  =t_{\pi}\sin^{2}\left(  \frac{a_{0}k_{x}}{2}\right)
+t_{\delta}\sin^{2}\left(  \frac{a_{0}k_{y}}{2}\right)  +t_{\pi}\sin
^{2}\left(  \frac{a_{0}k_{z}}{2}\right)  ,\nonumber\\
\varepsilon_{3}  &  =t_{\pi}\sin^{2}\left(  \frac{a_{0}k_{x}}{2}\right)
+t_{\pi}\sin^{2}\left(  \frac{a_{0}k_{y}}{2}\right)  +t_{\delta}\sin
^{2}\left(  \frac{a_{0}k_{z}}{2}\right)  , \label{enrs}%
\end{align}
where $a_{0}$ is the lattice constant. The matrix $W$ describes the mixing of
subbands within the conductivity band. For the cubic phase of SrTiO$_{3}$,
counting the band energy from the $G$ point (i. e., dropping a uniform shift
of the whole band), $W$ is given by:%
\begin{equation}
W=\left(
\begin{array}
[c]{ccc}%
0 & \xi & \xi\\
\xi & 0 & \xi\\
\xi & \xi & 0
\end{array}
\right)  . \label{W}%
\end{equation}
For the tetragonal phase as reported in Ref. \cite{VDM2011}, the matrix $W$
is:%
\begin{equation}
W=\left(
\begin{array}
[c]{ccc}%
2D & \xi & \xi\\
\xi & 2D & \xi\\
\xi & \xi & -4D
\end{array}
\right)  .
\end{equation}

\end{document}